\documentclass[%
 reprint,
superscriptaddress,
pra,
]{revtex4-2}
\usepackage{amsmath}
\usepackage{amsfonts}
\usepackage{amssymb}
\usepackage{float}

\usepackage[version=4]{mhchem}
\usepackage{stmaryrd}
\usepackage{graphicx}
\usepackage[export]{adjustbox}
\graphicspath{ {./images/} }

\usepackage{dcolumn}
\usepackage{bm}
\usepackage[colorlinks=true,linkcolor=blue,urlcolor=blue,citecolor=blue]{hyperref}
\usepackage[mathlines]{lineno}
\usepackage{xcolor}
 
\begin{document}

\preprint{APS/123-QED}

\title{
A comprehensive study of the Spin-Hall effect of tightly focused linearly polarized light through a stratified medium in optical tweezers
}
\author{Sramana Das}
\email{sd22rs032@iiserkol.ac.in}
\affiliation{Department of Physical Sciences, Indian Institute of Science Education and Research Kolkata, Mohanpur-741246, West Bengal, India} 

\author{Sauvik Roy}
\email{sr19rs022@iiserkol.ac.in}
\affiliation{Department of Physical Sciences, Indian Institute of Science Education and Research Kolkata, Mohanpur-741246, West Bengal, India}

\author{Subhasish Dutta Gupta}
\affiliation{Department of Physical Sciences, Indian Institute of Science Education and Research Kolkata, Mohanpur-741246, West Bengal, India}
\affiliation{School of Physics, Hyderabad Central University, India}
\affiliation{Tata Institute of Fundamental Research Hyderabad, India}

\author{Nirmalya Ghosh}
\email{nghosh@iiserkol.ac.in}
\affiliation{Department of Physical Sciences, Indian Institute of Science Education and Research Kolkata, Mohanpur-741246, West Bengal, India}

\author{Ayan Banerjee}
\email{ayan@iiserkol.ac.in}
\affiliation{Department of Physical Sciences, Indian Institute of Science Education and Research Kolkata, Mohanpur-741246, West Bengal, India}

\begin{abstract}
The optical Spin-Hall effect originates from the interaction between the spin angular momentum (SAM) and extrinsic orbital angular momentum (OAM) of light, leading to mutual interrelations between the polarization and trajectory of light in case of non-paraxial fields. Here, we extensively study the SHE and the resultant Spin-Hall shifts (SHS) in optical tweezers (OT) by varying the numerical aperture of objective lenses, and the refractive index (RI) stratification of the trapping medium. Indeed, we obtain much larger values of the SHS for particular combinations of NA and stratification compared to the sub-wavelength orders typically reported. We also observe that the longitudinal component of the spin angular momentum (SAM) density - which is responsible for the spin of birefringent particles in optical tweezers - changes more-or-less monotonically with the lens numerical aperture, except around values of the latter where the angle subtended by the focused light equals the critical angle for a particular RI interface. Our results may find applications in designing experiments for tuning the SHS and SAM induced due to SOI to generate exotic optomechanics of trapped particles in optical tweezers.
\end{abstract}

\maketitle


\section{\label{sec:level1}Introduction}

The angular momentum of light manifests itself in both spin (circular or elliptical polarization) and orbital (helical phase front) degrees of freedom. Though the spin angular momentum (SAM) and the orbital angular momentum (OAM) are independent quantities when light propagates paraxially through vacuum or isotropic homogeneous media~\cite{bliokh2014extraordinary,berry2009optical,barnett2016natures}, the picture does not remain the same in non paraxial cases. Thus, in processes when light propagates through inhomogeneous or anisotropic media,  is scattered by mesoscopic particles, or is tightly focused in isotropic inhomogeneous media, the Spin Orbit Interaction (SOI) of light couples the SAM (purely intrinsic) and OAM (intrinsic or extrinsic) -- leading to various fascinating optical phenomena \cite{o2002intrinsic,saha2018transverse,bliokh2015spin,liberman1992spin,marrucci2011spin}. These interactions and interconversions can be classified broadly into three categories according to the division of SAM and OAM in terms of their intrinsic and extrinsic nature: (i) Interaction between SAM and intrinsic OAM of light, leading to the generation of spin-induced optical vortices (ii) Interaction between SAM and extrinsic OAM, manifested as a spin-dependent shift of the trajectory of the light beam, popularly known as Spin-Hall effect of light\cite{onoda2004hall,leyder2007observation,hosten2008observation,yin2013photonic,qin2009measurement,ling2017recent}, and (iii) Interaction between intrinsic OAM and extrinsic OAM, generating intrinsic OAM-dependent shift of the beam trajectory --- known as the Orbital Hall effect \cite{bliokh2015quantum,korger2014observation,bliokh2006geometrical,merano2010orbital,zhang2014orbit}. Each of these effects have generated significant research interest and have potential applications in diverse areas involving light matter interactions \cite{roy2013controlled,roy2014manifestations}.

The optical Hall effect is essentially the transverse spatial separation of opposite angular momentum components, while conserving the total angular momentum of light . Accordingly, SHE is the transverse spin-dependent displacement of two opposite SAM density components, left circularly polarized and right circularly polarized light. The SHE of light is universal because the SOI exists in various scenarios  ranging from reflection or refraction of optical beams at interfaces between two media, scattering processes , tight focusing of fundamental and higher order Gaussian beams, high numerical aperture imaging geometry \cite{bliokh2015spin}, propagation through inhomogeneous anisotropic media \cite{bliokh2019spin} etc.
However, a pronounced SHE only arises due to breaking of cylindrical symmetry during spin-to-orbital angular momentum conversion \cite{zhao2007spin}, with the effect observed in the direction orthogonal to the symmetry-breaking axis. Importantly, most of the SOI effects originates due to the evolution of different geometric phases and their spatial and momentum gradients. In case of tight focusing by high NA lens in optical tweezers, the wavevectors of the incoming collimated beam are rotated in the meridional plane due to the lensing effect, and a conical $k$-distribution is generated in the focused field which induces a gradient in geometric phase. Now, the generation of a large longitudinal component of the electric field due to tight focusing, and the incorporation of a stratified media with large RI gradient in the path of light beam increase the geometric phase gradient of the focused light, leading to high SHS. Thus, with input linearly polarized light which is not spherically or cylindrically symmetric, the shift is expected to be enhanced by an appropriate choice of RI contrast of the stratified medium, and the degree of tight focusing of the beam which is determined by the NA of the objective lens.

Note that the SHE has been studied in stratified media using optical tweezers setup in the literature \cite{roy2014manifestations}. However, the role of NA of the objective lens, i.e. the effect of tight focusing in the SHE in presence of a stratified medium in the optical tweezers path has not been quantified yet. In this paper, we have addressed this aspect this problem in detail. We have numerically simulated the effect of NA and the RI gradient of a stratified medium on the SHE due to linearly polarized Gaussian beam. Our numerical simulation is built around the Debye-Wolf theory (or Debye-Wolf diffraction integrals) to quantify the fields near the focus in a four-layered stratified medium. Note that, while the spin Hall shift (SHS) is a phenomena which typically occurs at sub-wavelength length scale, here we obtain a much larger shift for certain RI stratification. Thus, we lay out a clear strategy to regulate the amount of shift, as well as the longitudinal component of the SAM density (which is a measure of the total longitudinal SAM that a birefringent probe particle may perceive on interacting with the field) of light by tuning the RI gradient of the stratified medium. Our study has important implications in experiments, since the enhanced spin-orbit conversion may lead to interesting applications in controlling the complex dynamics of optically trapped particles in optical tweezers \cite{roy2014manifestations}. In addition, our study also provides a recipe towards the use of a stratified medium as a further tool to control the SHS and SAM density in a systematic manner.

The structure of the paper is as follows. In Sec. II, we lay out the theoretical formulation needed to evaluate the field components near the focus in stratified medium. Then, in Sec.III, we discuss the numerical results in optical tweezers, and analyze the results in each case. Finally, we summarize and conclude our work in Sec.IV.

\section{\label{sec:level1}Theoretical Formulation}

The Debye-Wolf theory or Angular Spectrum method \cite{richards1959electromagnetic,wolf1959electromagnetic} is applied to analyze the electric field in the vicinity of the focal plane of an objective lens when linearly polarized light is coupled into it, also accounting for the influence of a stratified medium traversed by the light\cite{rohrbach2005stiffness,novotny2012principles,rodriguez2010optical}. This formalism for Gaussian beams has been extensively researched to address various experimental scenarios within optical tweezers configurations \cite{torok1997electromagnetic,haldar2012self}. This technique enables the determination of fields at a specific point of interest through the computation of two-dimensional vector diffraction integrals for all constituting plane waves known as spatial harmonics.

\begin{figure}[H] 
        \centering \includegraphics[width=0.45\textwidth]{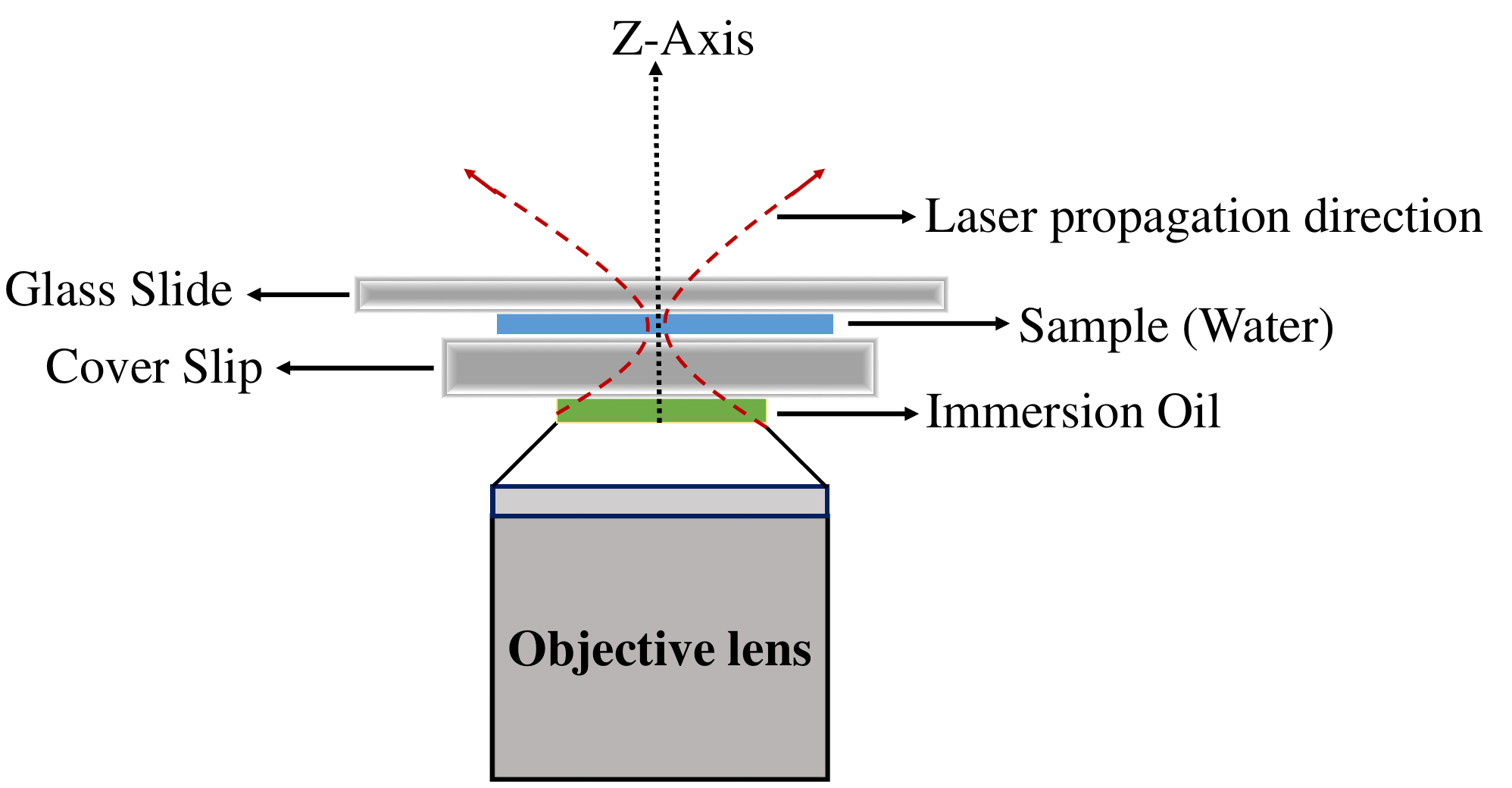}
        \caption{
                Schematic of the stratified medium in OT that has been used in our study
        }
        \label{fig:2}
\end{figure}

To model the stratified medium in the path of laser beam a hybrid transfer function is used which can lead to the desired field profiles anywhere inside or after the stratified medium \cite{rohrbach2005stiffness,roy2013controlled}. This transfer function contains generalized Fresnel's coefficients which take care of reflected and transmitted fields due to stratification. In our case, the SHE is studied by varying NA of objective lens. The NA of a lens determines how tightly the light is focused into the medium - the more the NA, the more tightly is the light focused. A stratified medium has been modeled with light being focused into the third layer i.e the sample chamber [Fig.~\ref{fig:2}]. The model thus consists of a lens system of arbitrary NA with its axis along z which coincides with the direction of stratification.

In Fig.~\ref{fig:2}, light is presented as a ray which denotes a single plane wave (with a particular propagation direction) propagating from one z plane to another (z is the beam propagation axis). In all the cases studied, the origin is fixed at the lab frame. Accordingly the interface positions are selected with respect to this origin so that the focal point lies within the sample chamber. The incident fields are considered as monochromatic with temporal factor exp(-i$\omega$t). 
\subsection{The Debye-Wolf Integral and modeling of stratified medium }

\begin{figure}[H] 
        \centering \includegraphics[width=0.45\textwidth]{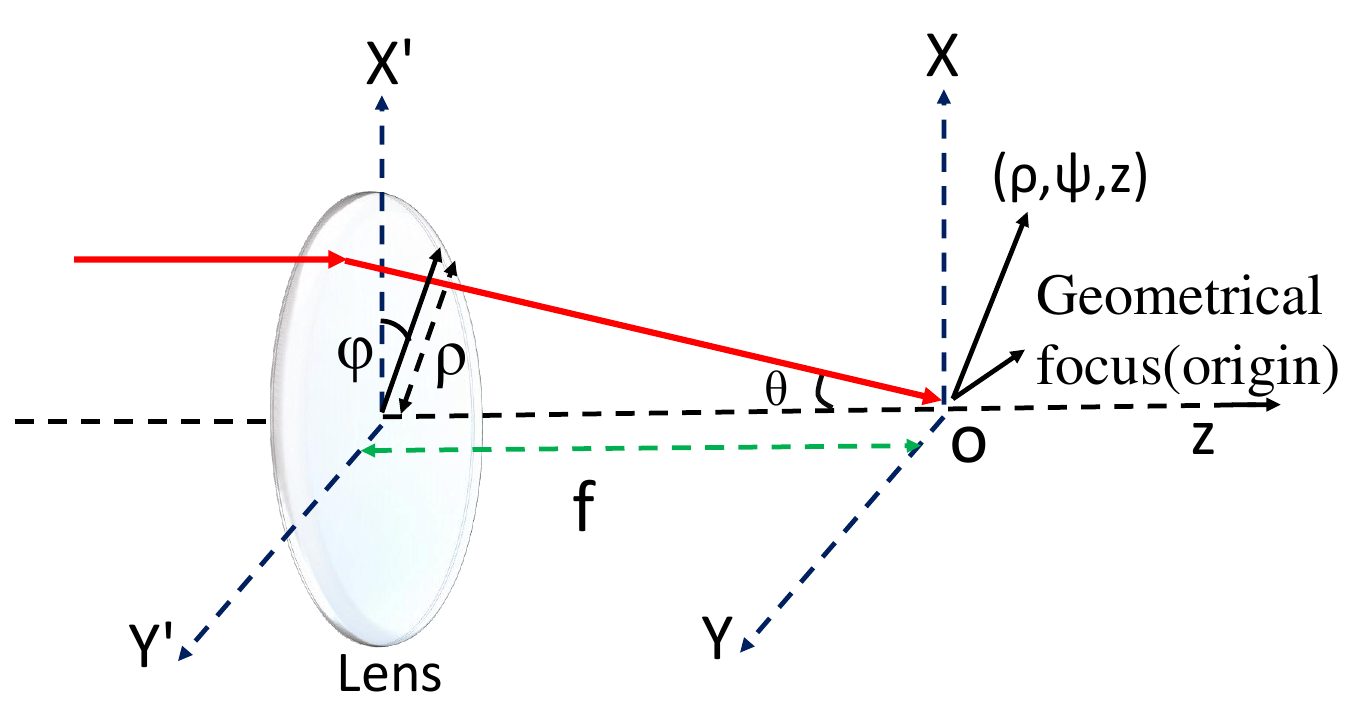}
        \caption{
                \label{fig:samplesetup} 
                Focusing by a lens (co-ordinate system used in field analysis)
        }
        \label{fig:3}
\end{figure}

The approach described in Ref.(\cite{richards1959electromagnetic,wolf1959electromagnetic}) - also known as the Angular Spectrum Method - is followed to produce the output fields near the beam focus. According to this formalism, the incident collimated Gaussian beam is decomposed into a superposition of plane waves having an infinite number of spatial harmonics. After focusing, the resultant field amplitude is related to incident field by a transfer function involving rotation matrices which imitates a lens action in laboratory frame. The co-ordinate frame that is used,  is shown in Fig.~\ref{fig:3}. 
As our system consists of 4-layered stratified medium, the polarization dependence of the fields needs to be considered. For that, the transfer function should incorporate the Fresnel transmission and reflection coefficients $T_s$, $T_p$ and $R_s$, $R_p$ to include the contribution of  multiple interfaces for $s$ and $p$ polarizations, respectively.

\vspace{1.5cm}
\begin{figure}[H] 
        \centering \includegraphics[width=0.45\textwidth]{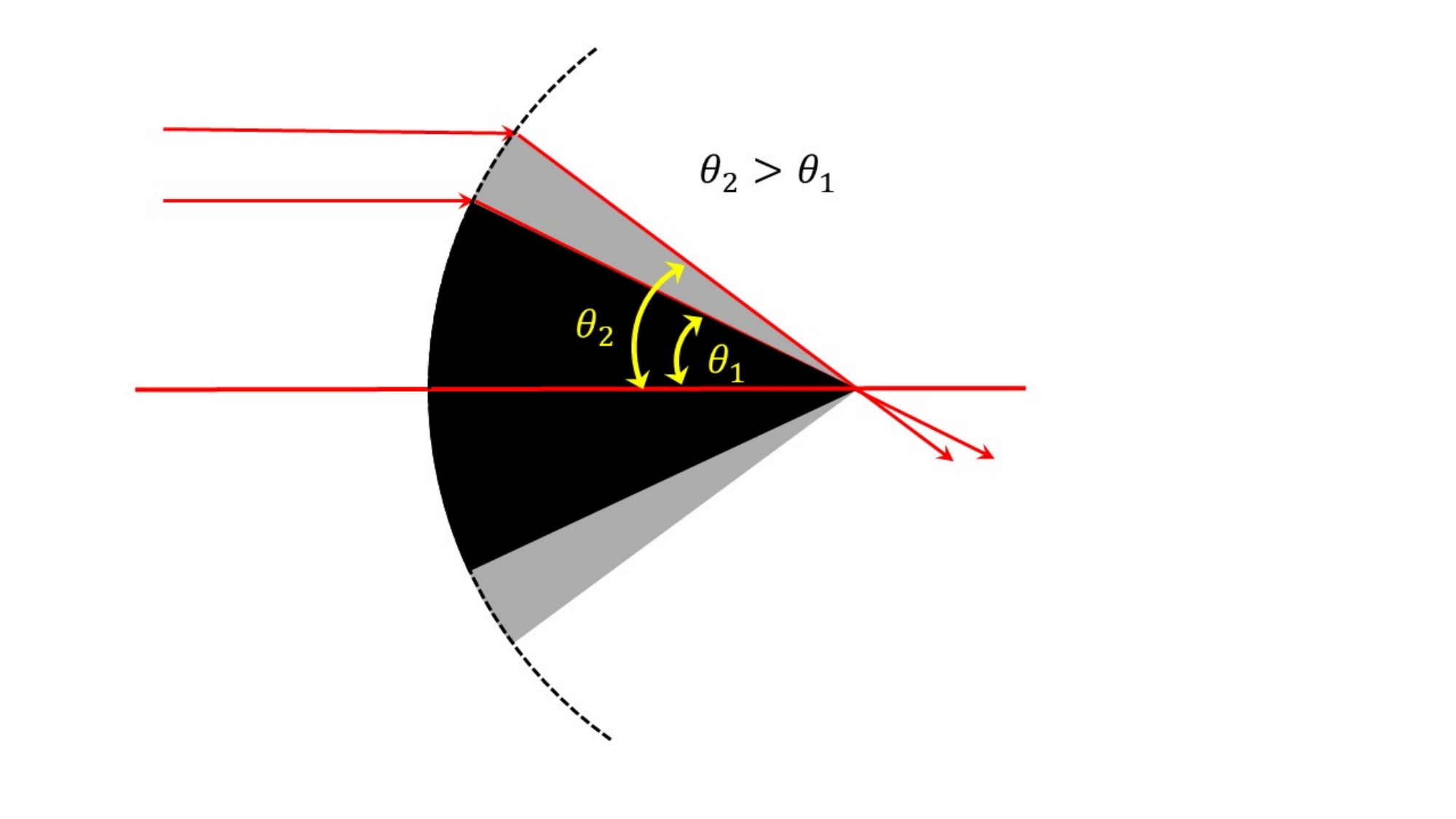}
        \caption{
                \label{fig:samplesetup} 
                NA of an objective lens portrayed as a series of infinitesimal solid angles.  
        }
        \label{33}
\end{figure}
Now, the output field amplitude $\vec{E}_{res}$($\theta$,$\phi$) is connected with input field amplitude by a modified transfer function as 
\begin{equation}
\vec{E}_{res}(\theta,\phi)=A\vec{E}_{inc}(\theta,\phi)    
\end{equation}

where $\vec{E}_{inc}$($\theta$,$\phi$) is the input field amplitude and  the modified transfer function,
\begin{equation}
A=R_z(-\phi) R_y(\theta) T R_z(\phi)  
\end{equation}
In general, $\vec{E}_{res}$($\theta$,$\phi$) is a superposition of forward and backward propagating waves in the stratified media, but the dominant contribution comes from the forward-propagating waves.

\vspace{0.3cm}
where,the SO(3) rotation matrices take the form as ,

\begin{equation}
R_{z}(\phi) = \begin{bmatrix}
\cos \phi & \sin \phi & 0 \\
-\sin \phi & \cos \phi & 0 \\
0 & 0 & 1
\end{bmatrix}
\end{equation}
\begin{equation}
R_{y}(\theta)  =\begin{bmatrix}
\cos \theta & 0 & -\sin \theta \\
0 & 1 & 0 \\
\sin \theta & 0 & \cos \theta
\end{bmatrix}
\end{equation}
The T matrix is given by,
\begin{equation}
T=\begin{bmatrix}
T_p & 0 & 0\\
0 & T_s & 0\\
0 & 0 & T_p
\end{bmatrix},  
\end{equation}

While this is for forward propagating waves, for backward propagating waves, we have to replace $T_s$ and $T_p$ by $R_s$ and $R_p$, respectively, in Eq. (5) and $\theta$ by $\pi$-$\theta$ in Eq. (2).

Now the resultant field can be obtained by integrating Eq.  (1) over $\theta$ and $\phi$ , so we finally get,

\begin{eqnarray*}
\vec{E}(\rho, \psi, z) & =i \frac{k f e^{-i k f}}{2 \pi} \int_{0}^{\theta_{\max }} \int_{0}^{2 \pi} \vec{E}_{\mathrm{res}}(\theta, \phi) e^{i k z \cos \theta} \\
& \times e^{i k \rho \sin \theta \cos (\phi-\psi)} \sin (\theta) d \theta d \phi, \hspace{1.7cm}(6)
\end{eqnarray*}

The limit for the $\theta$ integral is set by the NA of the objective lens $NA=n \sin\theta_{max}$, $n$ is the refractive index of the surrounding medium of objective lens.

The resultant field can also be written in matrix form for the linearly polarized light (x-polarized) as follows, 

\begin{eqnarray*}
{\left[\begin{array}{l}
E_{x} \\
E_{y} \\
E_{z}
\end{array}\right]=} & C\left[\begin{array}{ccc}
I_{0}+I_{2} \cos 2 \psi & I_{2} \sin 2 \psi & 2 i I_{1} \cos \psi \\
I_{2} \sin 2 \psi & I_{0}-I_{2} \cos 2 \psi & 2 i I_{1} \sin \psi \\
-2 i I_{1} \cos \psi & -2 i I_{1} \sin \psi & I_{0}+I_{2}
\end{array}\right] \\
& \times\left[\begin{array}{l}
1 \\
0 \\
0
\end{array}\right]=C\left[\begin{array}{c}
I_{0}+I_{2} \cos 2 \psi \\
I_{2} \sin 2 \psi \\
-i 2 I_{1} \cos \psi
\end{array}\right] .\hspace{1.7cm}(7)
\end{eqnarray*}

This is a general expression for both reflected and transmitted waves. $I_{0}$, $I_{1}$, and $I_{2}$ are the diffraction integrals given below

\begin{eqnarray*}
I_{0}^{t}(\rho)= & \int_{0}^{\min \left(\theta_{\max }, \theta_{c}\right)} E_{\mathrm{inc}}(\theta) \sqrt{\cos \theta}\left(T_{s}^{(1, j)}+T_{p}^{(1, j)} \cos \theta_{j}\right) \\
& \times J_{0}\left(k_{1} \rho \sin \theta\right) e^{i k_{j} z \cos \theta_{j}} \sin (\theta) d \theta \\
I_{1}^{t}(\rho)= & \int_{0}^{\min \left(\theta_{\max }, \theta_{c}\right)} E_{\mathrm{inc}}(\theta) \sqrt{\cos \theta} T_{p}^{(1, j)} \sin \theta_{j} J_{1}\left(k_{1} \rho \sin \theta\right) \\
& \times e^{i k_{j} z \cos \theta_{j}} \sin \theta d \theta \\
I_{2}^{t}(\rho)= & \int_{0}^{\min \left(\theta_{\max }, \theta_{c}\right)} E_{\text {inc }}(\theta) \sqrt{\cos \theta}\left(T_{s}^{(1, j)}-T_{p}^{(1, j)} \cos \theta_{j}\right) \\
& \times J_{2}\left(k_{1} \rho \sin \theta\right) e^{i k_{j} z \cos \theta_{j}} \sin \theta d \theta \hspace{1.7cm}(8)
\end{eqnarray*} \\
and
\begin{eqnarray*}
I_{0}^{r}(\rho)= & \int_{0}^{\min \left(\theta_{\max }, \theta_{c}\right)} E_{\text {inc }}(\theta) \sqrt{\cos \theta}\left(R_{s}^{(1, j)}-R_{p}^{(1, j)} \cos \theta_{j}\right) \\
& \times J_{0}\left(k_{1} \rho \sin \theta\right) e^{-i k_{j} z \cos \theta_{j}} \sin \theta d \theta \\
I_{1}^{r}(\rho)= & \int_{0}^{\min \left(\theta_{\max }, \theta_{c}\right)} E_{\mathrm{inc}}(\theta) \sqrt{\cos \theta} R_{p}^{(1, j)} \sin \theta_{k} J_{1}\left(k_{1} \rho \sin \theta\right) \\
& \times e^{-i k_{j} z \cos \theta_{j}} \sin \theta_{1} d \theta \\
I_{2}^{r}(\rho)= & \int_{0}^{\min \left(\theta_{\max }, \theta_{c}\right)} E_{\text {inc }}(\theta) \sqrt{\cos \theta}\left(R_{s}^{(1, j)}+R_{p}^{(1, j)} \cos \theta_{j}\right) \\
& \times J_{2}\left(k_{1} \rho \sin \theta\right) e^{-i k_{j} z \cos \theta_{j}} \sin \theta d \theta \hspace{1.7cm}(9)
\end{eqnarray*}
where, suffixes $t$ and $r$ denote the transmitted and reflected component, respectively. The $\phi$ integrals are related to Bessel functions $J_{n}$ and $I_{0}, I_{1}$, and $I_{2}$ play an important role in the resulting intensity distribution in the focal plane. Here, the cumulative sum of the constituent plane waves has been studied to analyse the focal fields following the angular spectrum method.

\subsection{Longitudinal Spin Angular Momentum}

Now, the distribution of Spin Angular momentum (SAM) density near focal plane is investigated after obtaining the electric fields in the previous section. The SAM density is defined by,
$$
\begin{gathered}
\mathbf{S}=\frac{1}{4 \omega} \operatorname{Im}\left[\varepsilon \mathbf{E}^{*} \times \mathbf{E}+\mu \mathbf{H}^{*} \times \mathbf{H}\right] = \mathbf{S}_{E}+\mathbf{S}_{H}\hspace{0.6cm}     (10)\\
\end{gathered}  
$$
where $S_{E}$ and $S_{H}$ are the electric and magnetic contributions respectively.
Our stratified medium is considered as nondispersive and nonmagnetic.In a lossless (refractive index n is real) and nonmagnetic medium,Eq.10 takes the form ,
$$
\begin{gathered}
\mathbf{S}=\frac{1}{4 \omega} \operatorname{Im}\left(\varepsilon_{0} n^{2} \mathbf{E}^{*} \times \mathbf{E}+\mu_{0} \mathbf{H}^{*} \times \mathbf{H}\right)\hspace{2.1cm}(11)
\end{gathered} 
$$
where $\varepsilon_{0}$ and $\mu_{0}$ are the permittivity and permeability of free space.

The z-component of SAM density i.e, the longitudinal component of S can be written as,
$$
\begin{gathered}
\mathbf{S}_{z}=\frac{1}{4 \omega}\left[\operatorname{Im}\left(E_{x}^{*} E_{y}-E_{x} E_{y}^{*}\right)+\operatorname{Im}\left(H_{x}^{*} H_{y}-H_{x} H_{y}^{*}\right)\right] \\
\end{gathered}
\\ 
\\ (12)
$$
where the subscripts x and y indicate the transverse directions. Further $S_z$ can be expressed as, 
$$
\begin{gathered}
S_z=2C\lvert E_x\rvert \lvert E_y \rvert Sin \phi\hspace{4.7cm}(13)
\end{gathered}
$$
where, C is a constant and $\phi$ is the phase difference between $E_x$ and $E_y$ components.
\section{\label{sec:level3}Results and Discussion}
\begin{figure*}[ht] 
        \centering \includegraphics[width=0.99\textwidth]{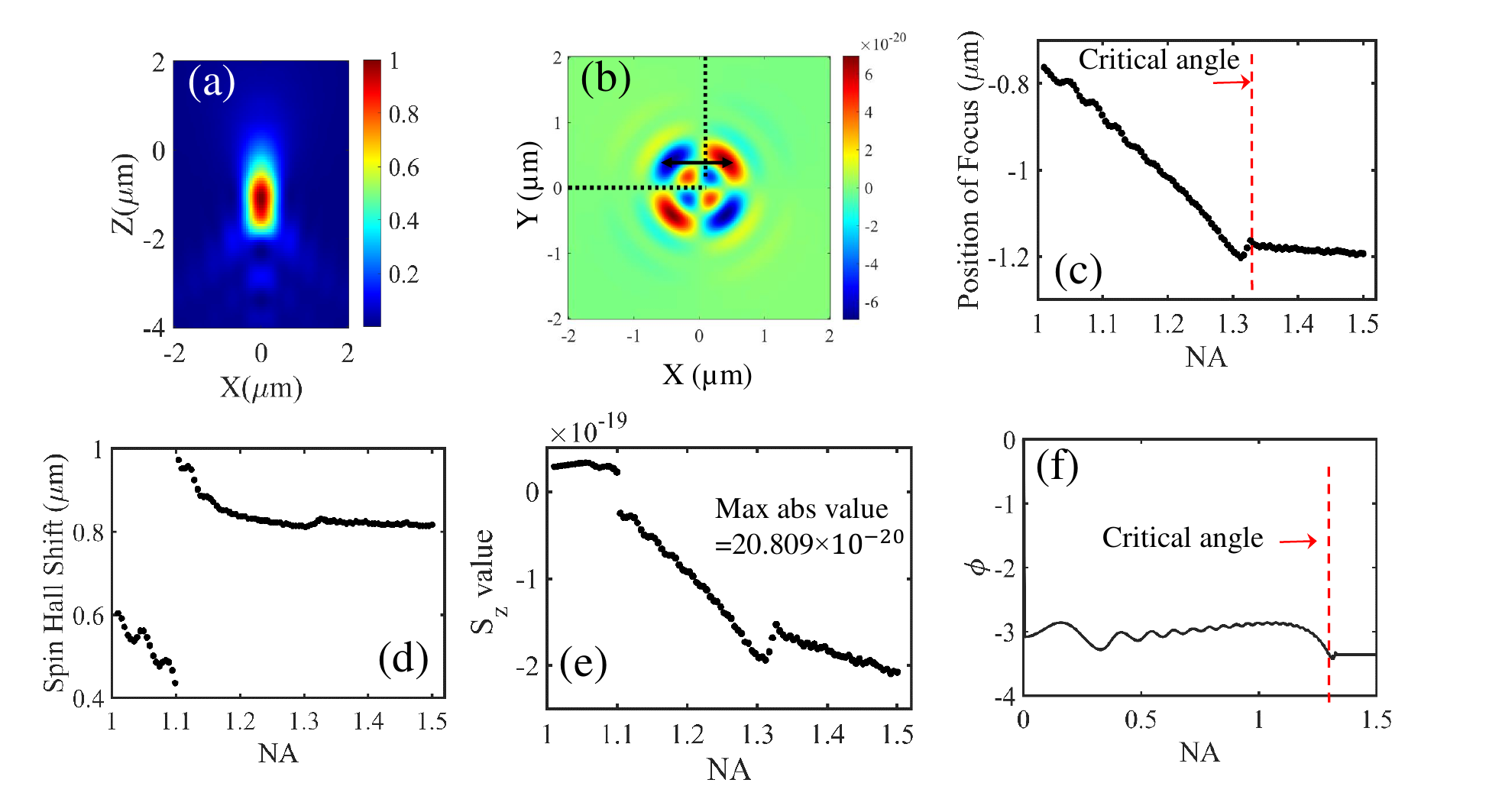}
        \caption{
               For the matched case, we show (a) $XZ$ plot of intensity for NA=1.5 (b) $S_z$ at NA=1.5. (c) the position of the focus changing with NA. (d) SHS. (e) Value of extrema of the $z$-component of SAM density taking electric part (S$_z$) considering the second quadrant. (f) Phase difference between $E_x$ and $E_y$ at an arbitrary point near the origin.}
               \label{fig:1}
\end{figure*}
\begin{figure*}[ht] 
        \centering \includegraphics[width=0.99\textwidth]{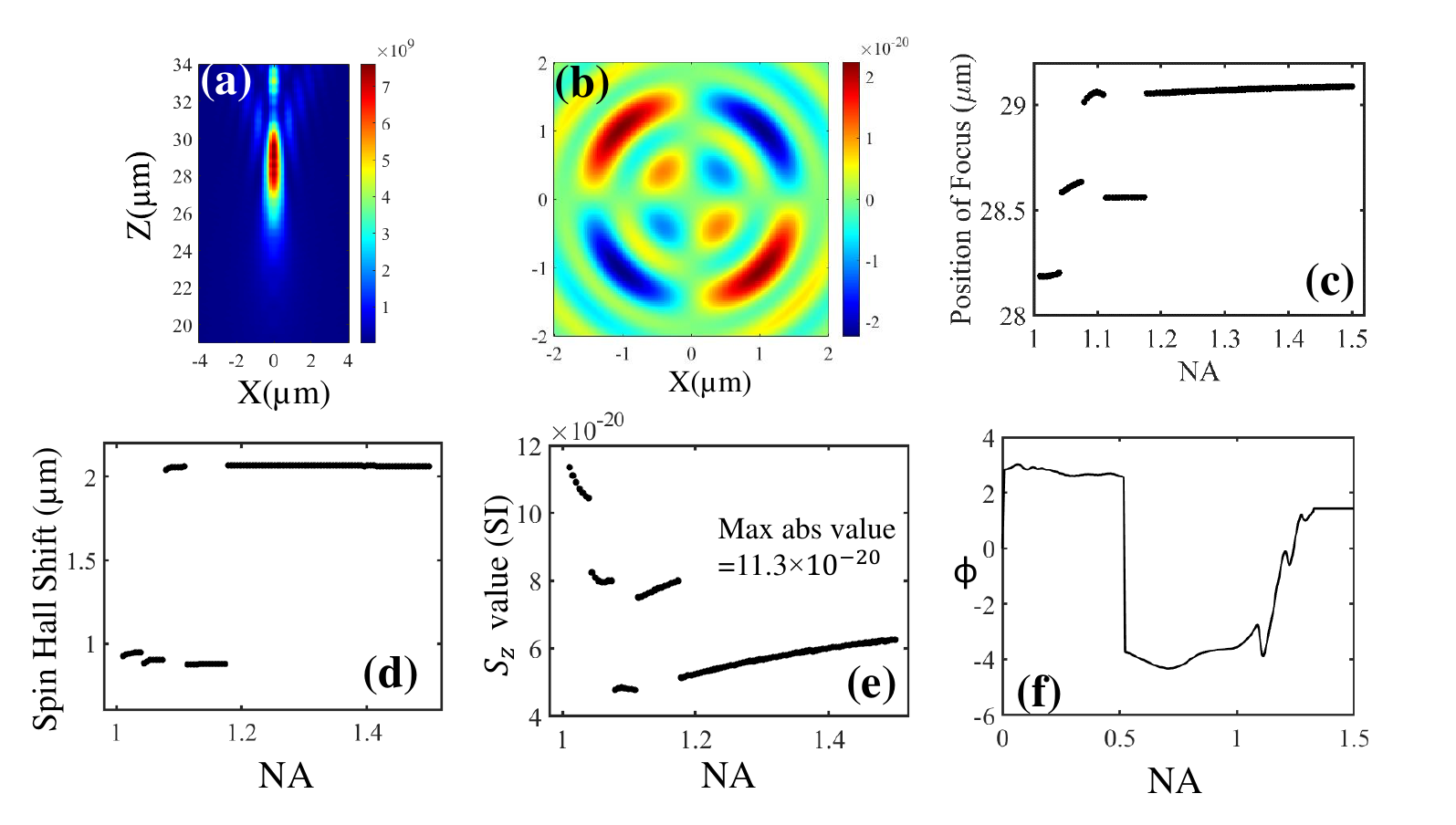} 
        \caption{
                \label{fig:samplesetup} 
               For the mismatched case, we show at 2$\mu$m away from focus (a) $XZ$ plot of intensity for NA=1.5 (b) $S_z$ at NA=1.5. (c) the position of the focus changing with NA. (d) SHS. (e) Value of extrema of the $z$-component of SAM density taking electric part (S$_z$) considering the second quadrant. (f) Phase difference between $E_x$ and $E_y$ at an arbitrary point near the origin.}
               \label{fig4}
\end{figure*}

The approach outlined above for generating the fields near the focus is implemented in a MATLAB script which is then utilized to analyze the focus position and the magnitude of the SHS as the NA of the objective lens increases. To encompass a broad selection of objectives typically employed for tight focusing, the investigated range of NAs (NA) spans from 1 to 1.5. An RI mismatched stratified medium containing multiple interfaces is another crucial element that is purposefully chosen to replicate a practical optical tweezers setup. This stratified medium contains four layers: 1) an immersion oil, 2) a coverslip, 3) the sample chamber, and 4) a top coverglass, acts as a modifier of the focused field. Two different coverslips -- one  matching the RI of the immersion oil (i.e., RI 1.516), referred to as 'matched case' and another with a mismatch (RI 1.814), referred to as 'mismatched case', are employed to alter the focused field in the sample chamber (RI 1.33). 
It is to be noted that the nominal focii of different lenses always coincide in the same point which is an inherent characteristic of the Debye-Wolf formalism. For the matched coverslip, the interface arrangement puts the actual focus within the sample chamber for the entire range of NA. However, for the mismatched coverslip with the same interface configuration, the incoming waves get focused in the fourth layer (top coverglass) i.e., at a larger distance due to the large RI contrast among the layers, thus loosening the tightness of focusing. A different interface arrangement, details of which are described below, 
is chosen to put the focus again in the third layer for all the NA considered. This positioning of the focus deep inside the third layer suggests ignoring the evanescent components in the third layer arising due to the total internal reflections from the coverslip-water interface. It is to be noted that the thicknesses of the coverslip of $160\mu m$ and the sample chamber of $15\mu m$ are kept the same in both the matched and mismatched cases.

The 4-layered stratified medium used in the MATLAB code for simulating the electric fields due to a linearly polarized Gaussian beam focused by an objective lens is shown in Fig.~\ref{fig:2}, with the RI arrangement described earlier. The linearly polarized Gaussian beam is derived from a laser of wavelength 1064 nm, and is first incident on an objective lens followed by the different interfaces of the stratified medium. In the present work, we have studied five distinct stratified media where the RI gradients are different. The positions of the three interfaces have been chosen with respect to the origin (z=0) set at laboratory co-ordinate frame in each case. The scenarios are : (a) Objective lens - immersion oil RI matched case: Here, the positions of the three interfaces with respect to that chosen origin are at -165 $\mu$m, -5 $\mu$m and 10 $\mu$m respectively. Here we vary NA from 1.01 to 1.5. (b) Objective lens - immersion oil RI mismatched case: Here, the positions of three interfaces are at -141 $\mu$m, 19 $\mu$m and 34 $\mu$m respectively. We studied two cases : one is at focus, and the other is at 2 $\mu$m away from focus (this axial distance is chosen to represent experimentally used scenarios \cite{roy2013controlled}. Once again, we vary the NA from 1.01 to 1.5. (c) Objective lens - air case : Here, the first layer is of air with RI=1 , not immersion oil. For this arrangement, too,  we studied two cases : one where refractive index of the coverslip is 1.516 and, another case where the refractive index of the same is 1.814. The positions of three interfaces are at -105 $\mu$m, 55 $\mu$m, 70 $\mu$m respectively for the first case and -90 $\mu$m, 70 $\mu$m, 85 $\mu$m, respectively, for the second. In both cases, we vary the NA from 0.5 to 1.

In our simulations, we studied three parameters: 1) position of the focus, 2) SHS, and 3) value of the $S_z$ (z-component of SAM density of light) in each of the above-mentioned cases, as a function of the objective lens NA, which basically determines how tightly the beam is focused into the third medium. Note that we have considered only the electric field contributions in all cases. Importantly, the intensity pattern in the transverse direction exhibited discrete lobe structures in general. The SHS was determined by the transverse separation between two intensity lobes having extremum value of opposite spin density [Fig.~\ref{fig:1}(b)]. For the $S_z$, we report only the extremum value of $S_z$ - which could lie either in the first, or second quadrant of the SAM density plot, depending on the NA. The plot for $S_z$ is symmetric, as a result of which any quadrant can be chosen. 

Now, we discuss the simulation results for each case based on the theoretical description and computational details mentioned above. The RI gradient of the stratified medium plays an important role that we extensively studied here. The stratified medium, understandably, modifies the size and shape of focal spot and the RI gradient determines the value of the SAM density, and SHS. As expected, we observe a correlation between value of $S_z$ and SHS with the RI gradient.
\subsection*{Study of Matched case}
We first showed the axial propagation of the light, where the diattenuation effects are not quite high, and off-axial spread in intensity is low [Fig. ~\ref{fig:1}(a)]. Fig.~\ref{fig:1}(b) shows an actual plot of $S_z$ for NA=1.5. Then we determined how the position of focus evolves with the NA of objective lens [Fig.~\ref{fig:1}(c)]. At first, the focal length decreases with NA - which is understandable, since higher NA implies tighter focusing. There is, of course, a distinct deviation from paraxial propagation, where the focal length should monotonically decrease with increase in NA, which we do not observe here.  There appears a small kink at a certain NA (NA=1.33), at which the maximum theta for the NA is equal to the critical angle for second and third medium. After that, the focal length is almost constant, since the the higher bending angles corresponding to the increased NA are beyond the critical angle corresponding to the interface, so that the rays incident at those angles do not propagate in the forward direction. 
\begin{figure*}[ht]
        \centering \includegraphics[width=0.99\textwidth]{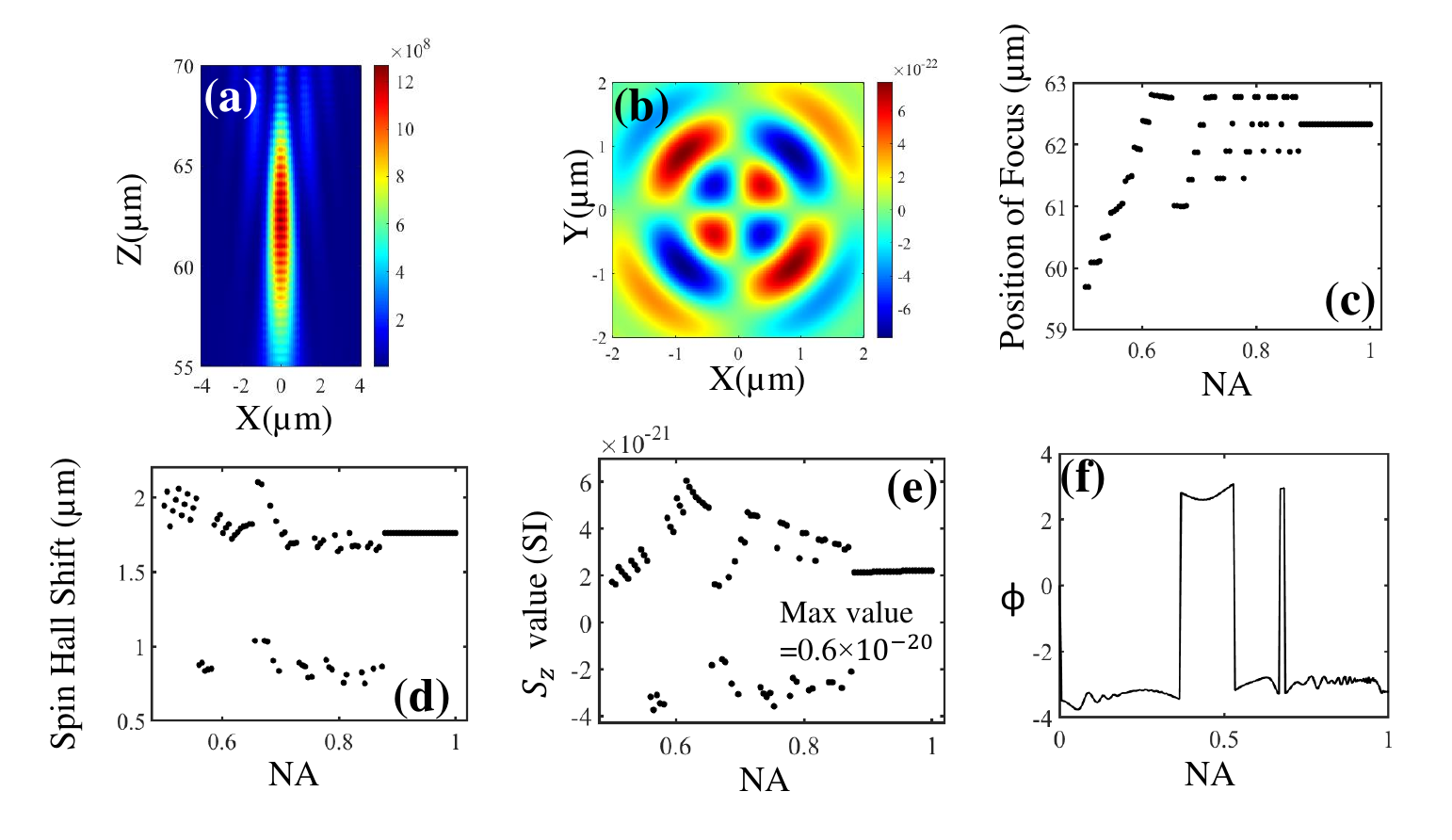}
        \caption{
                \label{fig:samplesetup} 
               In the air-Objective case (RI of the cover slip is 1.516), we show (a) Intensity in the $XZ$ plane for NA=0.9. (b) $S_z$ at NA=0.9. (c) the position of the  focus changing with NA. (d) SHS. (e) Value of extrema of z-component of SAM density taking electric part (S$_z$) considering the second quadrant (f) Phase difference between $E_x$ and $E_y$ at an arbitrary point near the origin.}
               \label{fig6}
\end{figure*}
\begin{figure*}[ht] 
        \centering \includegraphics[width=0.99\textwidth]{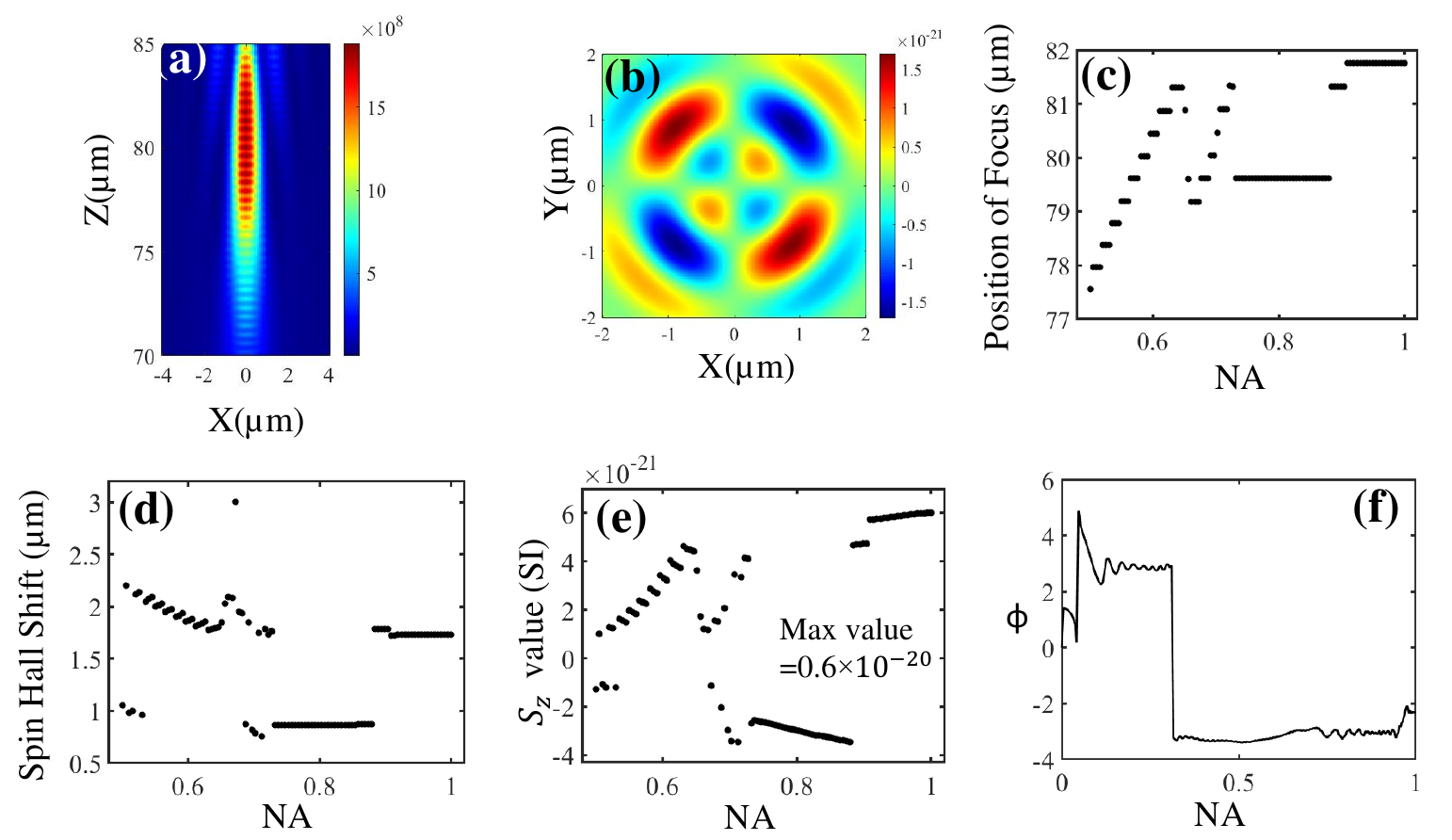}
        \caption{
                \label{fig:samplesetup} 
               In air-Objective case (RI of CoverSlip is 1.814), we show (a) XZ plot for focus at NA=0.9. (b) $S_z$ at NA=0.9. (c) the position of the focus changing with NA. (d) SHS. (e) Value of extrema of z-component of SAM density taking electric part (S$_z$) considering the second quadrant (f) phase difference between $E_x$ and $E_y$ at an arbitrary point near the origin.}
               \label{fig7}
\end{figure*}

 Note that we determine both the SHS, and the extremum of $S_z$ as a function of NA, which we show subsequently. Thus, the maximum SHS is around 1$\mu$m [(Fig.~\ref{fig:1}(d)], which occurs at NA=1.1, after which it reduces somewhat, and then becomes constant beyond an NA of 1.3. $S_z$, however, increases (in the negative direction) monotonically with NA [(Fig.~\ref{fig:1}(e)]. A kink is again observed at NA=1.33, as observed in the SHS plot as well. To understand this kink, we plot the phase difference between $E_x$ and $E_y$ (Fig.~\ref{fig:1}(f)) at an arbitrary point near the origin since there are finite values of the field at those regions. Here, we consider the entire range of solid angles for a particular NA to plot the phase difference (see Fig.~\ref{33}). Indeed, the phase plot shows the same behaviour as the SHS and $S_z$, with oscillations at certain NA values, and a jump at NA = 1.33. The phase is also indicative of a finite diattenuation parameter \cite{roy2013controlled}, which manifests itself as a non-zero field energy in the transverse direction (transverse components of the Poynting vector \cite{roy2013controlled}) due to interface effects. Beyond NA=1.33, $S_z$ increases again monotonically. 

\subsection{Study of Mismatched case}

In the RI mismatched case, the RI gradient is much higher, so that the behaviour of SHS and $S_z$ are rather complicated. The focal shift became virtually constant after NA=1.2 (see Appendix). Interestingly, in this case - the effects of spherical aberration are substantial \cite{roy2013controlled}, so that the focal spot size itself is large, as is the value of the diattenuation parameter \cite{roy2013controlled}, resulting in the formation of high-intensity rings in the transverse direction, as the light propagates axially. This has led to very interesting experimental observations and allow for diverse optomechanical effects, as we have shown earlier \cite{haldar2012self}. Therefore, we studied two cases here, z=0, and z = 2 $\mu$m away from focus. Here we show only the results for z = 2 $\mu$m (Fig.~\ref{fig4}), since this is the region which has produced interesting experimental results in the literature \cite{roy2013controlled,roy2014manifestations}. 

In  Fig.~\ref{fig4}(a), we show the axial propagation of the beam, which displays the enhanced transverse spread both beyond and before the focus, and even some refocusing effects at higher axial distances.
We then provide an actual plot of $S_z$ in Fig.~\ref{fig4}(b) where, distinct lobes of opposite helicity are observed at large transverse distances from the beam center. The focal position again is a function of the NA, as we show in  Fig.~\ref{fig4}(c). Clearly, in this scenario of RI-stratification, the SHS is larger than that in the matched case. Thus, the maximum shift is around 2 $\mu$m in Fig.~\ref{fig4}(d) which is close to 2 times the incident laser wavelength (1.064 $\mu$m). The SHS values also show discrete jumps at certain values of NA, and then becomes constant after an NA of 1.2 [Fig.~\ref{fig4}(d)]. Interestingly, the highest SHS occurs at at NA between 1.1 and 1.2. The maximum value of $S_z$ also displays discrete jumps, and increases gradually beyond NA of 1.2 (Fig.~\ref{fig4}(e). This behaviour is different from the matched case, as is the fact that the value of $S_z$ is less (11.3$\times10^{-20}$ versus 20.8$\times10^{-20}$). The trends in both SHS and $S_z$ are understandable from Fig.~\ref{fig4}(f), where we again plot the phase between the orthogonal electric field components. The oscillations are observed here as well, along with the fact that the phase differences here are also higher in magnitude compared to the matched case (Fig.~\ref{fig:1}(f) - which is actually the reason behind the enhanced diattenuation parameter here.  We intend to explore these effects for future experimental works in SOI-induced optomechanics in the near future. 

In both the matched and mismatched cases, we observe the SHS to be highest around a particular value of NA as is shown in Table \ref{11}, after which it decreases. As the NA is increased further, the SHS saturates when the corresponding maximum angle is beyond the critical angle, and we ignore the evanescent component. It is interesting, however, that the SHS maximizes at a value of NA that is not obvious. We attribute this to the fact that both geometric and dynamical phase appear in focusing in the presence of interfaces, with the diffraction integrals containing both terms. Now, SHS is only connected to geometric phase, since it is the latter which leads to the generation of the helicity components. It is thus independent of the dynamic phase, which is the same for the two helicity component, and thus cancels out in determining the SHS (which is the transverse separation between two opposite helicity components).  However, since the SHS depends on the geometric phase gradient, the maximization of the SHS depends on both the total geometric phase acquired by the light after focusing through the stratified medium, and the size of the focal spot so that it is apparently optimized at a particular NA value. Now, in case of $S_z$, both geometric and dynamic phase contribute, since it is given by the ratio of $I_2$/$I_0$ the phase difference between the orthogonal field components and thus increases monotonically with increasing NA. This is an important finding of our work. Note that the dependence of $S_z$ on NA and RI stratification is provided in Table \ref{22}.
\setlength{\arrayrulewidth}{0.5mm}
\setlength{\tabcolsep}{18pt}
\renewcommand{\arraystretch}{1.5}

\begin{table*}

\begin{tabular}{ |p{3cm}|p{4.5cm}|p{3.5cm}|  }
\hline
\textbf{RI-stratification}& \textbf{Maximum SHS (in $\mu$m)}    & \textbf{Corresponding NA} \\
\hline
1.516,1.516,1.33,1.516 (water) & 0.97 (at focus) & 1.1 \\
\hline
1.516,1.814,1.33,1.516 (water) & 1.28 (at focus)  & 1.11   \\
\hline
1.516,1.814,1.33,1.516 (water)& 2.0 (at 2$\mu$m away from focus)  & 1.2 to 1.5  \\
\hline
1,1.516,1.33,1.516 (air) & 2.1 (at focus) & 0.66 \\
\hline
1,1.814,1.33,1.516 (air)& 3.0  (at focus) & 0.67 \\

\hline

\end{tabular}
\caption{SHS as a function of NA and RI stratification}
\label{11}
\end{table*}

\vspace{1cm}

\setlength{\arrayrulewidth}{0.5mm}
\setlength{\tabcolsep}{18pt}
\renewcommand{\arraystretch}{1.5} 

\begin{table*}
\begin{tabular}{ |p{3cm}|p{4.5cm}|p{3.5cm}|  }
\hline
\textbf{RI-stratification} & \textbf{Maximum $S_z$ (in SI)}    & \textbf{Corresponding NA} \\
\hline
1.516,1.516,1.33,1.516 (water)& 20.809$\times10^{-20}$ (at focus) & 1.1 \\
\hline
1.516,1.814,1.33,1.516 (water) & 5.87$\times10^{-20}$ (at focus)  & 1.5   \\
\hline
1.516,1.814,1.33,1.516 (water)& 11.3$\times10^{-20}$ (at 2$\mu$m away from focus)  & 1.01  \\
\hline
1,1.516,1.33,1.516 (air)& 0.6032$\times10^{-20}$ (at focus) & 0.99 \\
\hline
1,1.814,1.33,1.516 (air)& 0.6025$\times10^{-20}$  (at focus) & 0.67 \\

\hline

\end{tabular}
\caption{ Absolute maximum values of $S_z$ as a function of NA and RI stratification}
\label{22}
\end{table*}

 \subsection{Study of Air-objective lens case}
\vspace{0.3cm}
We now consider the case where first layer of stratified medium is air instead of immersion oil, leading to much higher RI gradient than earlier. Once again, we studied two scenarios where the RI of the coverslips were different, viz. 1.516 and 1.814. Fig.~\ref{fig6} and Fig.~\ref{fig7} show our results for the case of RI 1.516 and 1.814 respectively. The beam shows signatures of refocusing as it travels axially (Fig.~\ref{fig6} and ~\ref{fig7}(a)) and in this case, we observe high values of the diattenuation parameter which is manifested as substantial transverse intensity distributions at various $z$ values as the beam propagates axially. Then we show the actual plot of $S_z$ for the both cases in Fig.~\ref{fig6} and ~\ref{fig7}(b). Quite interestingly, the focal shifts show an interesting periodicity here Fig.~\ref{fig6} and ~\ref{fig7}(c), with NA. Also, while the SHS values as shown in Fig ~\ref{fig6} and ~\ref{fig7}(d) are much higher than that in the earlier cases - leading to shifts of close to 2 and 3 times the incident laser wavelength (for RI of 1.516 and 1.814, respectively), the value of $S_z$ is more than an order of magnitude lower than the two earlier cases we considered Fig ~\ref{fig6} and ~\ref{fig7}(e). This is understandable, since the values of NA are much lower in air compared to oil immersion objectives. This shows the non-linear dependence of $S_z$ on NA, where a change of 50\% leads to much larger change in both geometric and dynamic phase. Once again, to understand the behaviour of SHS and $S_z$ we plot the phase difference between $E_x$ and $E_y$ (Fig.~\ref{fig6} and~\ref{fig7} (f)) at an arbitrary point near the origin. It is clear, therefore, that optical tweezers with air-based objectives would facilitate multi-axial trapping akin to holographic tweezers. 

\section{\label{sec:level3}Conclusion}
In this paper, we perform a thorough study of the effect of tight focusing (NA varying between 1-1.5 for immersion oil objectives, and 0.5-1 for air-based objectives) and RI gradient on SHS and $S_z$ in optical tweezers for input linearly polarized light.  Our results are rather intriguing where it is clear that the SHS increases with increasing RI gradient, but has a more complicated relationship with NA. Thus, for low RI gradient (matched case in immersion oil objectives), it maximizes at NA=1.1, but reduces below that and saturates beyond an NA of 1.33. For the mismatched case, the SHS once more is highest at NA = 1.1-1.2, but reduces thereafter. On the other hand, the absolute maximum value of $S_z$ decreases with increasing RI-gradient, but it generally increases with increasing NA, except at values where the focused beam is incident on an interface at critical angle, when the phase between the orthogonal components of the electric field undergoes a sudden change, leading to concomitant effects in both SHS and $S_z$. 
Indeed, we observe that this phase factor for input linearly polarized light entirely dictates the behaviour of both SHS and $S_z$, which is again understandable, since it affects both the diattenuation (responsible for SHS), and the ellipticity (controlling the value of $S_z$) of the electric field. The evolution of both SHS and $S_z$ as a function of NA are understandable in terms of their dependence on geometric and dynamic phases of light, with the former dependent only on the geometric phase gradient, and the latter on the geometric and dynamic phases acquired by light as it is focused into the stratified medium. Our work thus shows the relevance of both the NA of the objective lens and RI contrast as tools for controlling SOI effects in the tightly focused light fields existing in optical tweezers, which would lead to more sophisticated avenues for the implementation of SOI-induced optomechanics of trapped particles in optical tweezers. In the near future, we would like to perform experiments to validate our simulated findings of this paper. We hope to report new results soon.

\section*{Acknowledgements}
The work was supported by IISER Kolkata, an autonomous teaching and research institute supported by the
Ministry of Human Research Development, Government of India. SD is thankful to the University Grant Commission,
Government of India for financial support through the Junior Research Fellowship(JRF) grant.


%

\end{document}